\documentclass[aps,prl,twocolumn,showpacs,preprintnumbers,amsmath,amssymb]{revtex4}
\usepackage{graphicx}% Include figure files
\usepackage{dcolumn}% Align table columns on decimal point
\usepackage{bm}% bold math

\begin{document}

%\preprint{APS/123-QED}

\title{Semi-Markov models of mRNA-translation}% Force line breaks with \\
\author{Mieke Gorissen$^1$}
% \altaffiliation[Also at ]{Physics Department, XYZ University.}
\author{Carlo Vanderzande$^{1,2}$}
\affiliation{%
$^1$Faculty of Sciences, Hasselt University, 3590 Diepenbeek, Belgium. \\
$^2$Instituut Theoretische Fysica, Katholieke Universiteit Leuven, 3001 Heverlee, Belgium.
} 
%\\

%\author{Carlo Vanderzande}
%\affiliation{Departement WNI, Hasselt University, 3590 Diepenbeek, Belgium 
%\\
%Instituut Theoretische Fysica, Katholieke Universiteit Leuven, 3001 Heverlee, Belgium
%}%

\date{\today}% It is always \today, today,

\begin{abstract}
Translation is the cellular process in which ribosomes make proteins from information encoded on messenger RNA (mRNA).  We model translation with an exclusion process taking into account the experimentally determined, non-exponential, waiting time between steps of a ribosome. From numerical simulations using realistic parameter values, we determine the distribution $P(E)$ of the number of proteins $E$ produced by one mRNA. We find that for small $E$ this distribution is not geometric.  We present a simplified and analytically solvable semi-Markov model that relates $P(E)$ to the distributions of the times to produce the first $E$ proteins.  \end{abstract}
\maketitle
Biological cells respond to external or internal signals by producing proteins. According to the central dogma of molecular biology  \cite{Alberts08}, the production of proteins from genes occurs in two steps. In the first one,  called {\it transcription}, information encoded in a gene is read by an RNA-polymerase and used to synthesize a messenger RNA (mRNA) molecule. In the second step, named {\it translation}, a ribosome moves along the mRNA and uses the information stored on its codons to make a new protein \cite{Alberts08}. 

Cellular  processes are inherently stochastic \cite{Ray08} because the various molecules involved occur in small numbers \cite{McAdams97}.
In this Letter, we will focus on stochastic aspects of the translation process.
We also take into account that mRNA is an unstable molecule which, through the action of RNase, decays with a rate $\lambda$. For example, the mRNA that produces the protein tsr-Venus in {\it E. Coli} has an average lifetime $1/\lambda=90$ seconds \cite{web10,Yu06}. 

A simple theory for stochastic protein production was proposed  more than thirty years ago by O. Berg \cite{Berg78}.  
Assuming that proteins are produced with a rate $k_t$, he showed that the number of proteins produced by one mRNA follows a geometric distribution with average $k_t/\lambda$. Recently, it has become possible to measure this distribution experimentally \cite{Yu06, Cai06} and fair agreement with Berg's theory was found. Because of its simplicity, the geometric distribution is also used in theoretical models of stochastic gene expression \cite{Friedman06,Azaele09}. Yet, Berg's theory cannot be the complete story and in this Letter we examine the robustness of his results to the addition of a few more realistic ingredients.

Each mRNA molecule has two different ends labelled as $3'$ and $5'$. A ribosome first attaches to the $3'$ end ({\it initiation}) after which it moves forward codon by codon towards the $5'$ end ({\it elongation}). In each step, an amino acid is added to the growing protein. Upon reaching the $5'$ end, the ribosome detaches ({\it termination}) and the newly synthesized protein is released. Each of the processes of initiation, elongation and termination involves several biochemical steps \cite{Martin09}. These individual steps can be considered to be memoryless (Markov) and therefore to have an exponential waiting time distribution. But on a more coarse grained level, quantities like the waiting time between two elongation steps generically have more complicated distributions. Indeed, recent single molecule experiments have shown that this so called {\it dwell time} can be fitted well by a gamma-distribution or by a difference of exponentials \cite{Wen08}. 

Berg's theory also does not take into account that at a particular moment, several ribosomes are attached to a mRNA. A given ribosome can then only move forward if there is sufficient free space in front of it. This leads to delays and is an extra source of stochasticity \cite{Dobrzynski09}. A well known model that describes these effects  \cite{MacDonald69} is the totally asymmetric exclusion process (TASEP) with extended objects  \cite{MacDonald69, Shaw03}.  In the TASEP, mRNA is represented as a one-dimensional lattice of $L$ sites, each site corresponding with one codon, that  can be occupied by at most one ribosome. Ribosomes are large in comparison with a codon and are therefore modelled as an extended 'particle' that covers $l$ lattice sites.  Following \cite{Shaw03}, we will take $l=12$. In this Letter, we study a version of the TASEP with extended particles where the dwell time $\tau$  has a waiting time probability density (WTPD) $\psi(\tau)$ that is gamma-distributed
\begin{eqnarray}
\psi(\tau) = \frac{\tau^{n-1} k^n}{\Gamma(n)} \exp\left[-k\tau\right]
\label{1}
\end{eqnarray}
Stochastic processes of this type with non-exponential waiting times are called semi-Markov in the mathematical literature.
In our model, ribosomes attach to the $3'$-end of mRNA with rate $\alpha$ (provided the first $12$ sites are empty), leave the $5'$-end with rate $\beta$, and the whole translation process stops with rate $\lambda$ due to degradation of mRNA. These latter three process are still treated as Markovian. Fig. 1 shows the ingredients of our model.
\begin{figure}[here]
\includegraphics[width=8.0cm]{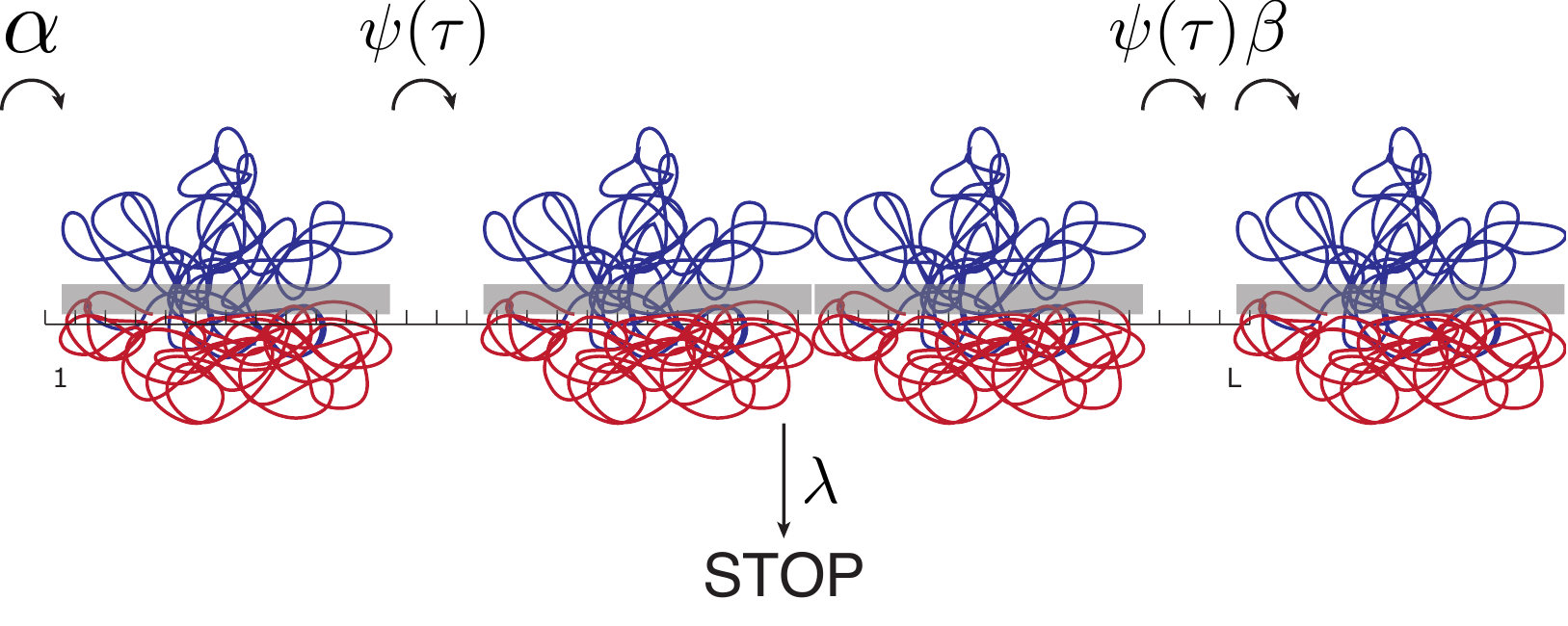}
\caption{\label{fig1} (Color online) 
Semi-Markov lattice model of translation. mRNA is modelled as a lattice of $L$ sites, each site representing one codon. Translation is initiated with rate $\alpha$ and terminated with rate $\beta$ while mRNA degrades with rate $\lambda$. Elongation proceeds with a waiting time probability density $\psi(\tau)$.}
\end{figure}

The average waiting time for the gamma-distribution (\ref{1}) is $n/k$, so that the ribosome on average needs a time $k/n$ to make one elongation step. We will take this as the unit of time in our simulations, so that we can put $k=n$ in (\ref{1}). For {\it E. Coli}, this corresponds in good approximation to $1/15$ seconds \cite{web10}. For the initiation and termination rate, we take $3$ respectively $1/2$ seconds, as determined experimentally \cite{Mitarai08}. In the chosen time unit, we therefore have $\alpha=1/45$ and $\beta=2/15$. Finally, we ran most of our simulation for $\lambda=1/1350$ and $L=815$, the values appropriate for the tsr-Venus protein. 

In order to simulate a semi-Markovian TASEP, we used a modified continuous time Monte Carlo method. In our algorithm, every ribosome has a clock associated to it, which states the remaining waiting time until the next attempt at movement. Once a clock reaches zero, the attempt is made, and the clock is (re)set. The waiting time distribution for each clock depends on the site the particle is currently located at: for the first site and last site, the waiting time is exponentially distributed with parameter $\alpha$ respectively $\beta$, while for sites in the bulk, we sample the waiting time from $\psi(\tau)$ as given in (\ref{1}). At the beginning of the simulation of a given realisation, we also draw a random time $\tau_0$ from an exponential distribution $\lambda e^{-\lambda t}$. Once the time of the simulation exceeds $\tau_0$, that history is stopped.

In the simulations, we start at $t=0$ with an empty lattice, and then initiate translation. We monitor the current $j_r(t)$ of ribosomes leaving the mRNA at the $5'$-end. {\it In the absence of decay}, the total number of proteins produced up to time $t$, $E(t)$, equals the time-integrated ribosome current $E(t)=\int_0^t j_r(t')dt'$. After some time, we expect our model to reach a non-equilibrium steady state (NESS) in which quantities like the average $\langle j_r(t)\rangle$ and the variance $\Delta j_r(t)$ of the current become time-independent. In Fig. 2, results for these quantities are shown. We also have indicated with a vertical line $t=5/\lambda$, a time at which almost all mRNA will have decayed. We see that the average and the variance of the number of proteins produced per time unit increases very rapidly before reaching a steady state value. We also notice that this time is much greater than the average lifetime $1/\lambda$ of mRNA. Ribosomes therefore seem to work in an early time regime where the current and its fluctuations are rather small. This may be a mechanism to reduce fluctuations in the number of proteins produced.
\begin{figure}[here]
\includegraphics[width=8.2cm]{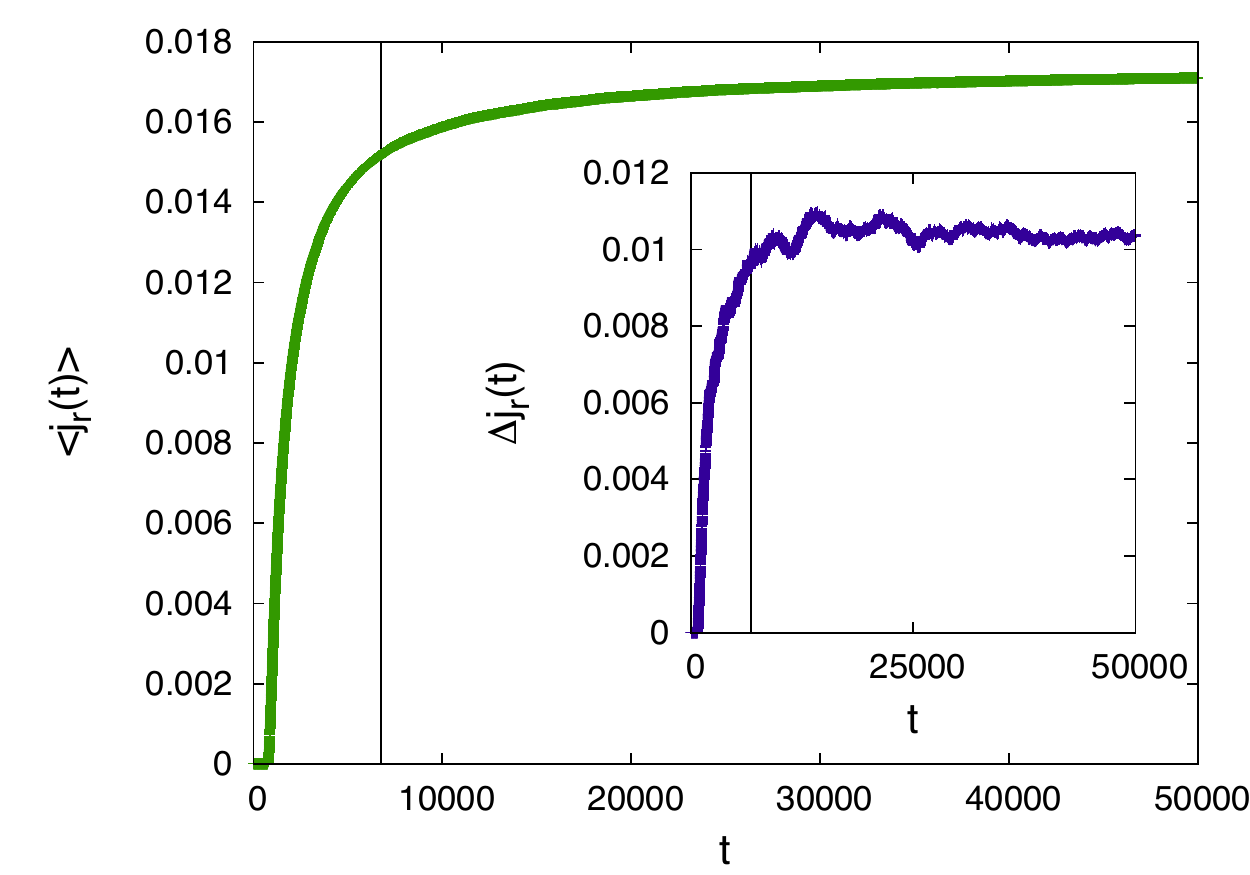}
\caption{\label{fig2} (Color online) 
Average and variance (inset) of the ribosome current leaving mRNA as a function of time, in the absence of mRNA decay. The vertical line is at $t=5/\lambda$. Data shown are for $k=n=0.5$ and are averaged over $10^3$ histories.}
\end{figure}

We now return to the model with mRNA-decay. 
From our simulations we have determined the full distribution $P(E)$ of the number $E$ of proteins produced. Our results are shown in a semi-log plot in Fig. 3. We see, that for $E \geq 4$, the distribution is still geometric but that there is a deviation at the smallest $E$ values. In Fig. 3 (inset) we have plotted the experimental data for tsr-Venus \cite{Yu06} in a similar way. Notice that while these are consistent with a geometric distribution, they also show an excess of cases where only a few proteins are produced. This could be an experimental signature of the effect seen in our model. It is interesting to remark that for larger values of the initiation and/or termination rate, the range over which the data deviate from a geometrical distribution becomes larger. Given that our values of $\alpha$ and $\beta$ are rather rough estimates, and that they may vary from protein to protein, it is quite possible that in other realistic situations a larger deviation from geometric behaviour can be observed.
\begin{figure}[here]
\includegraphics[width=8.0cm]{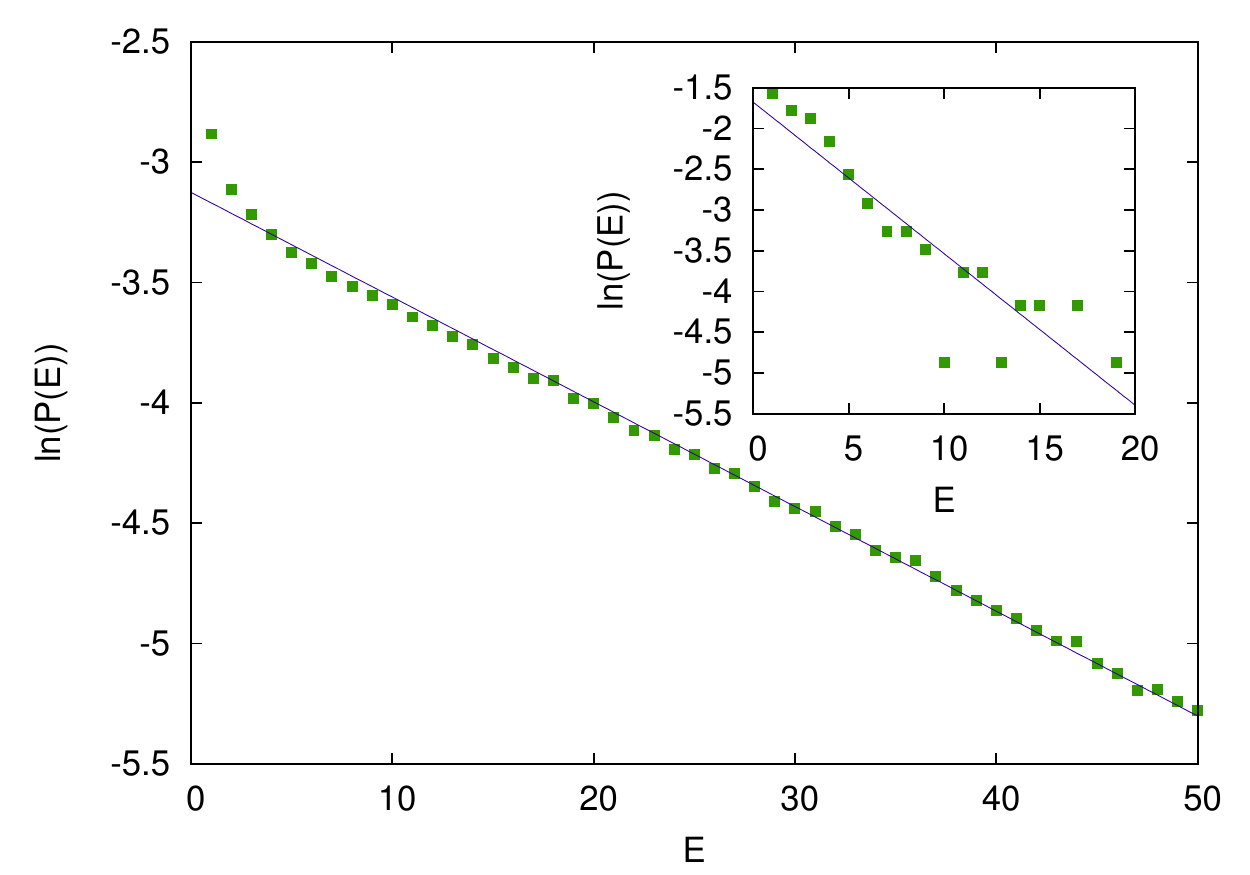}
\caption{\label{fig3} (Color online) 
Semi-log plot of the distribution $P(E)$ of the number of proteins produced as calculated using our model (for $k=0.5$) and from experiments (inset). The straight lines are best fits to a geometric distribution. }
\end{figure}

From our data, we have  calculated $\Delta E/\langle E\rangle^2$ (where $\Delta E$ and $\langle E\rangle$ are the variance and average of the number of proteins produced), a common noise measurement in the experimental literature on stochastic gene expression. This quantity was found to depend only weakly on $k$. An average over a range of $k$-values gives $\Delta E/\langle E\rangle^2=.98\pm0.02$ to be compared with the experimental data in \cite{Yu06}, from which one finds $\Delta E/\langle E\rangle^2=.74$. Given the relative simplicity of our model and the fact that {\it no parameters were fitted} we judge the agreement to be quite satisfying. Adding extra detail to our model, like fast and slow bonds \cite{Mitarai08}, might improve the agreement.

Finally, we calculated the waiting time $\tau_E$ between the production of the $(E-1)$-th and $E$-th protein, given that mRNA did not decay.  Fig. 4 shows the probability densities $\psi_E$ of these waiting times for $E=1$ (inset) and $E=2,\cdots,6$. It takes a relatively long time to produce the first protein, but then protein production is on average periodic \cite{Estimate}. A more detailed look shows that the distributions become independent of $E$ from $E=4$ onwards. Moreover, they are clearly non-exponential. We were not able to fit these WTPD's with a simple probability density like that of the gamma-distribution.
\begin{figure}[here]
\includegraphics[width=8.0cm]{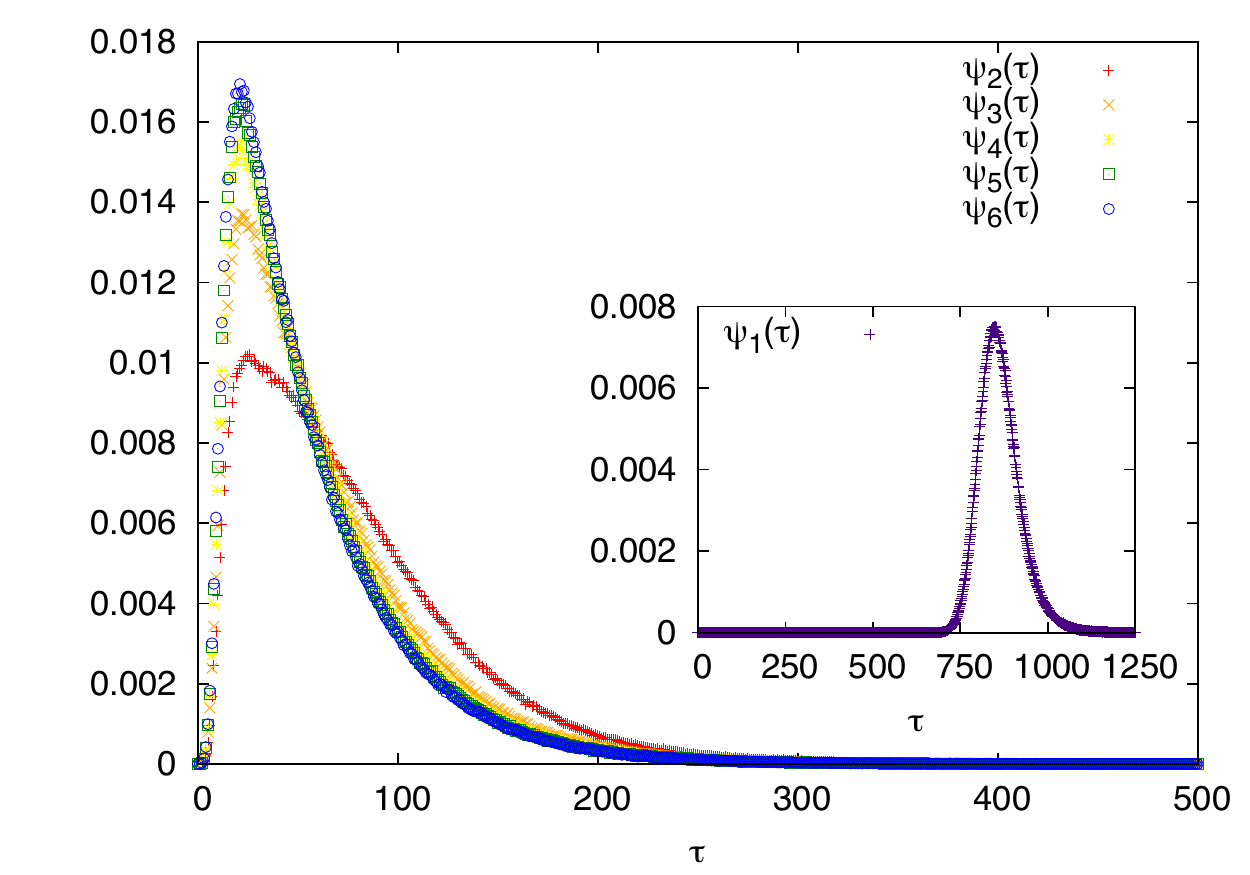}
\caption{\label{fig4} (Color online) 
Probability density $\psi_E$ of the waiting time between the production of the $(E-1)$-th and $E$-th protein for $E=1,\cdots,6$. Data are for $k=n=0.5$ and are averaged over $5\times10^6$ histories. }
\end{figure}

One of the interesting features of Berg's original model is its simplicity, which makes it very easy to use in calculations of the distribution of the number of proteins in a cell \cite{Friedman06,Azaele09}. Our lattice model is too complicated for this. We would therefore like to replace it by a simpler effective model in the spirit of Berg, yet sharing most of the properties of the full model shown in Fig. 1. This can be achieved by replacing Berg's reaction scheme for mRNA (Fig. 5, leftside) with a semi-Markov one (Fig. 5, rightside) in which mRNA decays with WTPD $\psi_\emptyset(\tau)$ and where the $E$-th protein is produced with WTPD $\psi_E(\tau)$. 

Conservation of probability implies
\begin{eqnarray}
\int_0^\infty\left[\psi_E(\tau)+\psi_\emptyset(\tau)\right]d\tau=1 \hspace{1cm} \forall E
\label{3}
\end{eqnarray}
We denote by $p_E=\int_0^\infty \psi_E(\tau)d\tau$, the probability that the $E$th protein is made before the mRNA has decayed.
\begin{figure}[here]
\includegraphics[width=8.0cm]{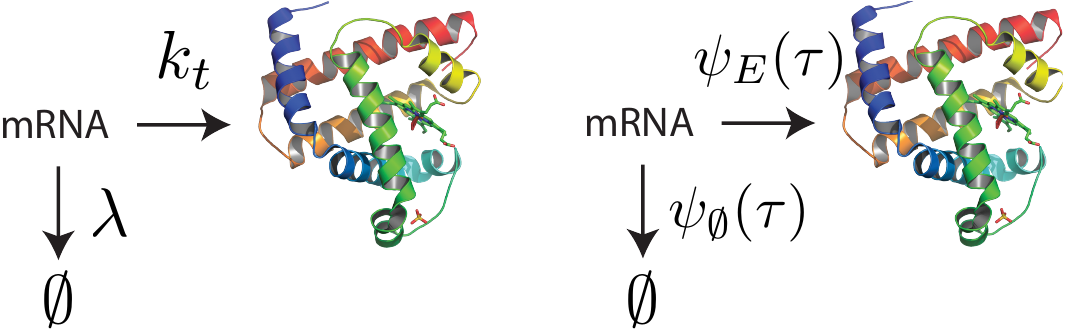}
\caption{\label{fig5} (Color online) 
Berg's standard model for protein production (left) and our semi-Markov extension (right). }
\end{figure}

Within this effective model, it is possible to calculate the distribution of the number of proteins produced exactly. We outline this derivation here. The state space of our semi-Markov model consists of the empty state $\emptyset$ representing a decayed mRNA, and the states $0, 1, 2, \cdots$ giving the number $E$ of proteins produced. We assume that at $t=0$, the process starts at $E=0$. In that case the probability $R(E,t)$  that the system is in the state $E$ at time $t$ evolves according to the generalised master equation \cite{Gillespie77,Esposito08}
\begin{eqnarray}
\frac{dR(E,t)}{dt}=\int_0^t K_{E}(t-\tau) R(E-1,\tau) d\tau \nonumber \\- \int_0^t \left[K_{E+1}(t-\tau) + K_\emptyset(t-\tau)\right] R(E,\tau)d\tau 
\label{4}
\end{eqnarray}
where the memory kernels $K_E$ and $K_\emptyset$ are defined in (\ref{6}) below and $R(-1,t)=0$. To solve this set of equations, we first do a Laplace transformations $\overline{R}(E,s)=\int_0^\infty e^{-st} R(E,t)dt$. Inserting the initial condition, (\ref{4}) becomes
\begin{eqnarray}
s\overline{R}(E,s) - \delta_{E,0}= \overline{K}_{E}(s) \overline{R}(E-1,s) \nonumber \\ - \left[\overline{K}_{E+1}(s) + \overline{K}_\emptyset (s)\right] \overline{R}(E,s)
\label{5}
\end{eqnarray}
where the Laplace transforms of the kernels can be related to those of the waiting time distributions \cite{Esposito08}
\begin{eqnarray}
\overline{K}_E(s)=\frac{\overline{\psi}_E(s)}{\overline{\phi}_E(s)} \hspace{1cm} \overline{K}_\emptyset(s)=\frac{\overline{\psi}_\emptyset(s)}{\overline{\phi}_E(s)}
\label{6}
\end{eqnarray}
and 
\begin{eqnarray}
\overline{\phi}_E(s)=\frac{1}{s}\left[1-\overline{\psi}_E(s)-\overline{\psi}_\emptyset(s)\right]
\label{7}
\end{eqnarray}
is the Laplace transform of $\phi_E(\tau)=1-\int_0^\tau \left[\psi_E(t)+\psi_\emptyset(t)\right]dt$, the probability that no transition from the state $E-1$ has been made up to time $\tau$.
Since $R(-1,s)=0$, the set of equations (\ref{5}) can be solved recursively giving
\begin{eqnarray}
\overline{R}(E,s)= \delta_{E,0} \overline{\phi}_{1}(s) + \frac{\overline{\psi}_{E}(s)\overline{\phi}_{E+1}(s)}{\overline{\phi}_E(s)} \overline{R}(E-1,s)
\label{8}
\end{eqnarray}

We are interested in the probability $P(E)$ that $E$ proteins are produced before the mRNA decays. This probability equals the probability that $E$ proteins are produced up to time $t'$ {\it times} the conditional probability that the mRNA decays in the time interval $(t',t'+dt')$ given that it didn't decay before, {\it summed} over all $t'$
\begin{eqnarray}
P(E) = \int_0^\infty R(E,t') \frac{\psi_\emptyset(t')}{\int_{t'}^{\infty} \psi_\emptyset(y)dy} dt'
\label{9}
\end{eqnarray}
We now assume, as in the lattice model, that mRNA-decay can still be described as a Markovian process with rate $\lambda$.  In that case, (\ref{9}) simplifies considerably and becomes
\begin{eqnarray} 
P(E) = \lambda \int_0^\infty R(E,t')dt' = \lambda \overline{R}(E,0)
\end{eqnarray}
so that $P(E)$ can be obtained from iterating (\ref{8}) at $s=0$. The iteration involves $\overline{\psi}_E(0)=p_E$ and $\phi_E(0)$ which using (\ref{3}) and (\ref{6}) becomes
\begin{eqnarray}
\overline{\phi}_E(0)=-\overline{\psi}_E'(0)-\overline{\psi}_\emptyset'(0)=\langle \tau_{E-1}\rangle
\end{eqnarray}
where $\langle \tau_{E-1}\rangle$ is the average time that the model stays in the state $E-1$. Putting everything together we find finally
\begin{eqnarray}
P(0)&=&\lambda \langle \tau_0\rangle \nonumber \\
P(E>0) &=& \lambda \left(\prod_{i=1}^E p_i\right) \langle \tau_{E}\rangle
\end{eqnarray}
Suppose now, that as was observed in our lattice model, $\psi_E(\tau)$ becomes independent of $E$ for $E\geq E_0$. We then find for $E\geq E_0$
\begin{eqnarray}
P(E)= \left[\lambda \left(\prod_i^{E_0-1} \frac{p_i}{p_{E_0}}\right) \langle \tau_{E_0}\rangle\right] (p_{E_0})^E
\label{12}
\end{eqnarray}
which is precisely a geometric distribution. In this way we understand that the observed deviations from the geometric distribution are related to the $E$-dependence of the WTPD $\psi_E(\tau)$. The simplified model therefore gives an explanation of the behaviour found in the lattice model.

In summary, we have studied the influence of memory effects on mRNA-translation. Taking as an input the experimentally determined waiting time distribution between two steps of a ribosome, we determine the distribution of the number of proteins produced, another experimentally accesible quantity. This distribution was found to deviate from a geometrical one. We could explain this behavior in terms of the properties of waiting time probabilities. 

The deviations from geometrical behavior become more important when the initiation and/or termination rates become larger than those found in living systems.
Recent progress has made it possible to study protein synthesis  in a cell-free system \cite{Karzbrun11}. It can be envisaged that such an approach can test the predictions following from our model by changing, for example initiation rates.

{\bf Acknowledgement} We would like to thank B. Wynants for useful discussions on semi-Markov processes. 


\begin{thebibliography}{100}
\bibitem{Estimate}
In the Markovian version of our model and in the absence of decay  the density of ribosomes as a function of $\alpha$ and $\beta$ can be calculated \cite{Shaw03}. From this density, one can estimate that the average time between the production of two proteins for our parameter values is $56$. This is close to the value found in the present model.
\end{thebibliography}
\end{document}